*Astro2010 State of the Profession Position Paper (March 2009)*

# Astroinformatics: A 21st Century Approach to Astronomy

**Primary Author:** Kirk D. Borne, Dept. of Computational and Data Sciences, 4400 University Drive MS 6A2, George Mason University, Fairfax, VA 22030 USA (kborne@gmu.edu).

Abstract:

Data volumes from multiple sky surveys have grown from gigabytes into terabytes during the past decade, and will grow from terabytes into tens (or hundreds) of petabytes in the next decade. This exponential growth of new data both enables and challenges effective astronomical research, requiring new approaches. Thus far, astronomy has tended to address these challenges in an informal and ad hoc manner, with the necessary special expertise being assigned to e-Science or survey science. However, we see an even wider scope and therefore promote a broader vision of this data-driven revolution in astronomical research. For astronomy to effectively cope with and reap the maximum scientific return from existing and future large sky surveys, facilities, and data-producing projects, we need our own information science specialists. We therefore recommend the formal creation, recognition, and support of a major new discipline, which we call Astroinformatics. Astroinformatics includes a set of naturally-related specialties including data organization, data description, astronomical classification taxonomies, astronomical concept ontologies, data mining, machine learning, visualization, and astrostatistics. By virtue of its new stature, we propose that astronomy now needs to integrate Astroinformatics as a formal sub-discipline within agency funding plans, university departments, research programs, graduate training, and undergraduate education. Now is the time for the recognition of Astroinformatics as an essential methodology of astronomical research. The future of astronomy depends on it.

## Preamble

New modes of discovery are enabled by the growth of data and computational resources in the sciences. This cyberinfrastructure includes databases, virtual observatories (distributed data), high-performance computing (clusters and petascale machines), distributed computing (the Grid, the Cloud, and peer-to-peer networks), intelligent search and discovery tools, and innovative visualization environments. Data volumes from multiple sky surveys have grown from gigabytes into terabytes during the past decade, and will grow from terabytes into tens (or hundreds) of petabytes in the next decade. This plethora of new data both enables and challenges effective astronomical research, requiring new approaches. Thus far, astronomy has tended to address these challenges in an informal and *ad hoc* manner, with the necessary special expertise being assigned to e-Science [24] or survey science. However, we see an even wider scope and therefore promote a broader vision of this data-driven revolution in astronomical research. The solutions to many of the problems posed by massive astronomical databases exist within disciplines that are far removed from astronomy, whose practitioners don't normally interface with astronomy. For astronomy to effectively cope with and reap the maximum scientific return from existing and future large sky surveys, facilities, and data-producing projects, we need our own information science specialists. We therefore recommend the formal creation, recognition, and support of a major new discipline, which we call ***Astroinformatics***. Astroinformatics **includes a set of naturally-related specialties including data organization, data description, astronomical classification taxonomies, astronomical concept ontologies, data mining, visualization, and statistics** [26]. By virtue of its new stature, we propose that astronomy now needs to integrate Astroinformatics as a formal sub-discipline within agency funding plans, university departments, research programs, graduate training, and undergraduate education. *Now is the time for the recognition of Astroinformatics as an essential methodology of astronomical research*. The future of astronomy depends on it.

## The Revolution in Astronomy and Other Sciences

The development of models to describe and understand scientific phenomena has historically proceeded at a pace driven by new data. The more we know, the more we are driven to enhance or to change our models, thereby advancing scientific understanding. This data-driven modeling and discovery linkage has entered a new paradigm [1], as illustrated in the accompanying graphic [2]. The emerging confluence of new technologies and approaches to science has produced a new Data-Sensor-Computing-Model synergism. This has been driven by numerous developments, including the information explosion, the development of dynamic intelligent sensor networks [http://www.thinkingtelescopes.lanl.gov/], the acceleration in high performance computing (HPC) power, and advances in algorithms, models, and theories. Among these, the most extreme is the growth in new data. The acquisition of data in all scientific disciplines is rapidly accelerating and causing a nearly insurmountable data avalanche [3]. Computing power doubles every 18 months (Moore's Law), corresponding to a factor of 100 in ten years. The I/O bandwidth (into and out of our systems, including data systems) increases by 10% each year – a factor 3 in ten years. By comparison, data volumes appear to double every year (a factor of 1,000 in ten years). Consequently, as growth in data volume accelerates, especially in the natural sciences (where funding certainly does not grow commensurate with data volumes), we will fall further and further behind in our ability to access, analyze, assimilate, and assemble knowledge from our data collections – unless we develop and apply increasingly more powerful algorithms, methodologies, and approaches. This requires a new generation of scientists and technologists trained in the discipline of data science [4].

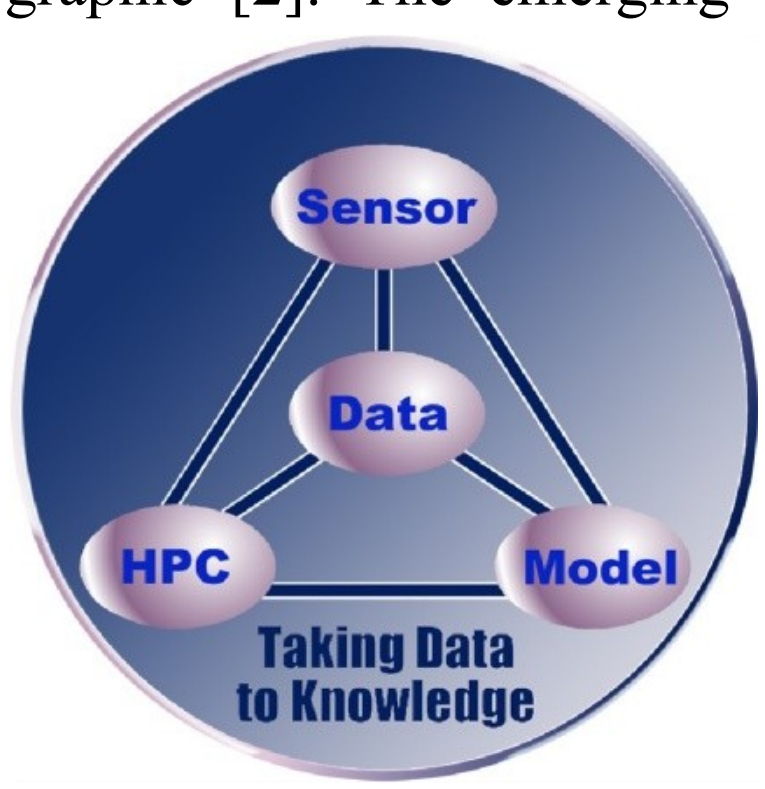

In astronomy in particular, rapid advances in three technology areas (telescopes, detectors, and computation) have continued unabated [5], all leading to more data [6]. With this accelerating advance in data generation capabilities over the coming years, we will require an increasingly skilled workforce in the areas of computational and data sciences in order to confront these challenges. Such skills are more critical than ever since modern science, which has always been data-driven, will become even more data-intensive in the coming decade [6, 7]. Increasingly sophisticated computational and data science approaches will be required to discover the wealth of new scientific knowledge hidden within these new massive scientific data collections [8, 9].

The growth of data volumes in nearly all scientific disciplines, business sectors, and federal agencies is reaching historic proportions. It has been said that *"while data doubles every year, useful information seems to be decreasing"* [10], and *"there is a growing gap between the generation of data and our understanding of it"* [11]. In an information society with an increasingly knowledge-based economy, it is imperative that the workforce of today and especially tomorrow be equipped to understand data and to apply methods for effective data usage. Required understandings include knowing how to access, retrieve, interpret, analyze, mine, and integrate data from disparate sources. In the sciences, the scale of data-capturing capabilities grows at least as fast as the underlying microprocessor-based measurement system [12]. For example, in astronomy, the fast growth in CCD detector size and sensitivity has seen the average dataset size of a typical large astronomy sky survey project grow from hundreds of gigabytes 10 years ago (e.g., the MACHO survey), to tens of terabytes today (e.g., 2MASS and Sloan Digital Sky Survey [5]), up to a projected size of tens of petabytes 10 years from now (e.g., LSST, the Large Synoptic Survey Telescope [6]). In survey astronomy, LSST will produce one 56Kx56K (3-Gigapixel) image of the sky every 20 seconds, generating nearly 30 TB of data daily for 10 years. In solar physics, NASA announced in 2008 a science data center specifically for the Solar Dynamics Observatory, which will obtain one 4Kx4K image every 10 seconds, generating one TB of data per day. NASA recognizes that previous approaches to scientific data management and analysis will simply not work. We see the data flood in all sciences (e.g., numerical simulations, high-energy physics, bioinformatics, drug discovery, medical research, geosciences, climate monitoring and modeling) and outside of the sciences (e.g., banking, healthcare, homeland security, retail marketing, e-mail). The application of data mining, knowledge discovery, and e-discovery tools to these growing data repositories is essential to the success of our social, financial, medical, government, and scientific enterprises. **An informatics approach is required.** What is informatics? Informatics has recently been defined as *"the use of digital data, information, and related services for research and knowledge generation"* [13], which complements the usual definition: *informatics is the discipline of organizing, accessing, integrating, and mining data from multiple sources for discovery and decision support* [14].

## A National Imperative

Our science education programs have always included the principles of evidence-based reasoning, fact-based induction, and data-oriented science [15]. In this age of the data flood, greater emphasis on and enhancement of such data science competencies is now imperative. In particular, we must muster educational resources to train a skilled data-savvy workforce: one that knows how to find facts (i.e., data, or evidence), access them, assess them, organize them, synthesize them, look at them critically, mine them, and analyze them.

The *Nature* article *"Agencies Join Forces to Share Data"* calls for more training in data skills [16]. This article describes a new Interagency Working Group on Digital Data representing 22 federal agencies in the U.S., including the NSF, NASA, DOE, and more. The group plans to set up a robust public infrastructure so that all researchers have a permanent home for their data. One option is to create a national network of online data repositories funded by the government and staffed by dedicated computing and data science professionals with science discipline expertise. Who will these computing and archiving professionals be? They will be a professional workforce trained in the disciplines of computational and data sciences and who collaborate with computer science and statistics professionals in these areas, including machine learning, visualization, statistics, algorithm design, efficient data structures, scalable architectures, effective programming techniques, information retrieval methods, and data query languages.

Within the scientific domain, data science is becoming a recognized academic discipline. F. J. Smith argues that now is the time for data science curricula in undergraduate education [17]. Others promote data science as a rigorous academic discipline [18]. Another states that *"without the productivity of new disciplines based on data, we cannot solve important problems of the world"* [19]. The 2007 NSF workshop on data repositories included a track on data-centric scholarship – the workshop report explicitly states our key message: *"Data-driven science is becoming a new scientific paradigm – ranking with theory, experimentation, and computational science"* [20]. Consequently, astronomy and other scientific disciplines are developing sub-disciplines that are information-rich and data-intensive to such an extent that these are now becoming (or have already become) recognized stand-alone research disciplines and full-fledged academic programs on their own merits. The latter include bioinformatics and geoinformatics, but will soon include astroinformatics, health informatics, and data science.

**National Study Groups Face the Data Flood**

Several national study groups have issued reports on the urgency of establishing scientific and educational programs to face the data flood challenges:

1. NAS report: *"Bits of Power: Issues in Global Access to Scientific Data"* (1997) [21];
2. NSF report: *"Knowledge Lost in Information: Report of the NSF Workshop on Research Directions for Digital Libraries"* (2003) [22];
3. NSB (National Science Board) report: *"Long-lived Digital Data Collections: Enabling Research and Education in the 21st Century"* (2005);
4. NSF report with the Computing Research Association: *"Cyberinfrastructure for Education and Learning for the Future: A Vision and Research Agenda"* (2005);
5. NSF "Atkins Report" : *"Revolutionizing Science and Engineering Through Cyberinfrastructure: Report of the National Science Foundation Blue-Ribbon Advisory Panel on Cyberinfrastructure"* (2005) [23];
6. NSF report: *"The Role of Academic Libraries in the Digital Data Universe"* (2006) [24];
7. NSF report: *"Cyberinfrastructure Vision for 21st Century Discovery"* (2007) [25];
8. JISC/NSF Workshop on Data-Driven Science & Repositories (2007) [20].

Each of these reports has issued a call to action in response to the data avalanche in science, engineering, and the global scholarly environment. For example, the NAS *"Bits of Power"* report lists five major recommendations, one of which includes: *"Improve science education in the area of scientific data management"* [21]. The Atkins NSF Report stated that skills in digital libraries, metadata standards, digital classification, and data mining are critical [23]. In particular, that report states: *"The importance of data in science and engineering continues on a path of exponential growth; some even assert that the leading science driver of high-end computing will soon be data rather than processing cycles. Thus it is crucial to provide major new resources for handling and understanding data."* [23] The core and most basic resource is the human expert, trained in key data science skills. As stated in the 2003 NSF "Knowledge Lost in Information" report, human cognition and human capabilities are fundamental to successful leveraging of cyberinfrastructure, digital libraries, and national data resources [22].

**E- Science in Astronomy**

A Library of Congress study group on e-Science [26] is examining the rapid growth in digital content, particularly in the sciences, with a goal to understand how the scholarly research environment is changing. The group leader, Dr. Peter Young, comments on the cyber-enabled data-driven revolution in the sciences in this way: *"These technologies are changing the conduct of science"* [private communication]. This data-driven transformation in the conduct of science has reached a level of maturity in the fields of bioinformatics and geoinformatics, recognized as stand-alone sub-disciplines of bio and geo, with their own funding programs, conferences, journals, and university departments. We propose that now is the time for similar recognition of Astroinformatics, at least with regard to its integration within agency funding plans, university departments, research programs, graduate training, and undergraduate education. **We define Astroinformatics as the formalization of data-intensive astronomy and astrophysics for research and education** [27, 28]. There is already a large and growing body of published research in this area, and we advocate programs to advance Astroinformatics as a viable formal sub-discipline of astronomy research and education. One of the core areas of Astroinformatics research is scientific data mining in astronomy [27]. A count of publications in this area (available through NASA's ADS) finds nearly 800 examples. If we change the search criteria to "neural networks", we find an additional 400 refereed papers. When we focus on astrostatistics, specifically looking at publications that applied one of the most common techniques (Bayesian analysis), we find another 2300 abstracts, of which nearly 600 are refereed papers. All of this clearly demonstrates an active productive research community in this field, many of whom have signed this document and who thereby support the formal recognition and development of an astronomical information sciences research discipline, which we label Astroinformatics.

The VAO (Virtual Astronomy Observatory, formerly NVO; hereafter VO) does not address this need because the VO addresses a completely different set of requirements. The VO is a grand information technology research e-Science program, whose goal has been to provide standards that describe all astronomical information resources worldwide, and to enable standardized discovery and access to these collections [29]. The emergence of VO has led to this new branch of astrophysics research – Astroinformatics – still in its infancy, consequently requiring further research and development as a discipline in order to aid in the data-intensive astronomical science that is emerging [27, 28, 30]. The VO and other astronomical community structures (i.e., ADASS, WGAS, and FITS WG) have been essential but incomplete steps toward an astronomical info/data sciences discipline area. There is still a need for data-intensive science research tools that mine and discover new knowledge from the vast distributed data repositories [27, 28]. One of the key challenges is that the various astronomical catalogs, databases, and observation logs have enormous variety in schema, metadata, information content, and knowledge representation. In most instances, the data are high-dimensional, thus requiring efficient and effective approaches (algorithms and data structures) for addressing the "curse of dimensionality" and for managing, mining, visualizing, and analyzing high-dimension data sets. Similarly, advances are required in the image sciences: computer vision, pattern recognition, visual information representation and retrieval, fusing images with different resolutions [31], and mining features in time series of images. Astroinformatics enables the integration and mining of these heterogeneous data/information/knowledge resources for scientific discovery.

**A Vision for Astroinformatics – The New Paradigm for Data-Intensive Astronomy**

Astroinformatics is **Discovery Informatics** for astronomy. We believe that it can and should become a standalone research discipline. Agresti [32] defined ***Discovery Informatics*** as: *"the study and practice of employing the full spectrum of computing and analytical science and technology to the singular pursuit of discovering new information by identifying and validating patterns in data."* The late Jim Gray (of Microsoft Research) championed the development of this fourth leg of science (data-intensive science), which he called **X-Informatics** (in KDD-2003), to accompany theory, computing, and experiment (observation). For our discipline, X is "astronomy". An informatics paradigm is needed within any data-intensive scientific discipline to make the best use of its rich data collections for scientific discovery. Discovery Informatics thereby activates data integration and fusion across multiple heterogeneous data collections to enable scientific knowledge discovery and decision support. Knowledge discovery is the central theme of science, and knowledge discovery in databases (KDD = data mining) is the "killer app" for scientific databases. KDD is an essential tool for 21st century research, because data-intensive science is here to stay (at petabyte scales and beyond). Discovering the "unknown unknowns" in enormous databases will lead to breakthroughs in scientific understanding in many fields. Discovery Informatics is therefore a key enabler for new science discovery in large databases, through the application of common data integration, browse, and discovery tools within a discipline (e.g., GIS tools in the field of Geoinformatics; BLAST, FASTA, and GeneOntology [GO] in Bioinformatics; QSAR in Cheminformatics). These common tools enable knowledge discovery to keep pace with exponentially growing data collections. The experience of the other data-intensive sciences reveals that is imperative for the successful pursuit of data-intensive astronomy that astronomers work with computer scientists to design, develop, and deploy a similar set of common tools, data models, and ontologies. The development of Astroinformatics as a research discipline focuses on these developments.

Astroinformatics includes: data models, data transformation and normalization methods, indexing techniques, information retrieval and integration methods, knowledge discovery methods, content-based and context-based information representations, consensus semantic annotation tags, taxonomies, ontologies, and more. These enable data mining, information retrieval and fusion, and knowledge discovery from huge astronomy datasets. But they do more – they also enable collaborative research and data re-use (both in the research environment and in learning settings). Astroinformatics provides a natural context for the integration of research and education – the excitement and experience of research and discovery are enabled and infused within the classroom through a portable informatics paradigm.

The database research community is responding to the data avalanche in astronomy, geosciences, and other disciplines [33]. Researchers have started an international effort to create an open-source database technology called SciDB that is designed fundamentally to address extreme-scale science [34]. SciDB's design is inspired primarily by the needs of the LSST, with guidance from other disciplines. The resulting architecture will enable analytics and knowledge generation on a scale that is practically unattainable with existing systems. Community support will ensure that SciDB and similar technologies will be available to meet the needs of data-intensive science.

The 2009 Semantic Astronomy workshop included a community discussion of the future of semantic research in astronomy. The consensus was that this community should organize around the single guiding principle of Astroinformatics [practicalastroinformatics.org], which means a focus on the mining and curation of information and knowledge derived from astronomical data.

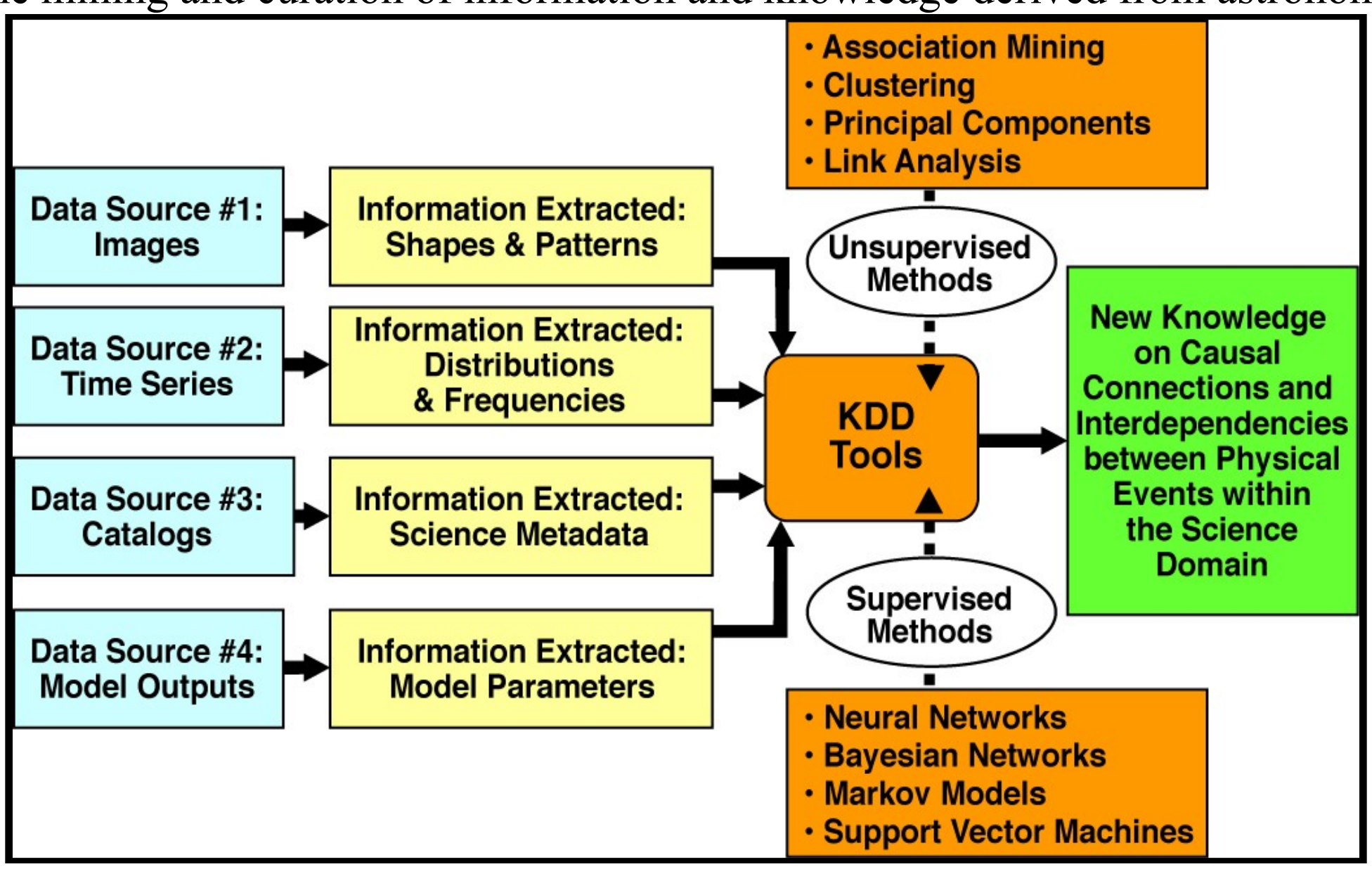

**Figure 2:** X-Informatics enables the KDD process flow for scientific knowledge discovery. The **informatics layer** is indicated by the boxes labeled "Information Extracted".

Figure 2 is drawn from the traditional DIKU (Data-to-Information-to-Knowledge-to-Understanding) flow from data to meaning (from sensors to sense) specifically within a scientific context [28]. This illustrates the KDD (Knowledge Discovery in Databases; i.e., data mining) process flow for scientific knowledge discovery: (a) from distributed heterogeneous data sources, we extract information of different types; (b) from the information, we mine new knowledge nuggets using a variety of KDD (Machine Learning) algorithms; and (c) with this knowledge, we apply our reasoning to attain greater understanding of the Universe. What is most interesting in Figure 2 is the informatics layer. On the left is the "data layer" – standardization is not required nor is it feasible, in this layer. Astronomy has excellent discipline-wide agreement on data syntax (the standard data format FITS), but patchy agreement on context and semantics: how the data are stored, organized, indexed, queried, or presented to the user; and how to repurpose and independently interpret the data. The middle layer is the **"informatics layer"** – this is where standardized representations of the "information extracted" are needed, for use in the "KDD layer" (data mining layer). This informatics layer includes standardized scientific metadata, taxonomies, ontologies, UCDs (Uniform Content Descriptors), astronomical vocabularies, and XML-based (self-documenting) representations of the information content extracted from the data [27, 28]. We observe from this diagram that the informatics representations of the extracted information are necessarily discipline-specific.

### Some Astroinformatics Research Use Cases

As illustrated in Figure 2, Astroinformatics enables data integration, data mining, and knowledge discovery across heterogeneous massive data collections. In addition, Astroinformatics enables many other use cases: re-use (and re-purposing) of archival data for new projects, semantic integration of data within different contexts (e.g., follow-up observations with robotic telescopes, collaborative research environments, search engines), literature-data linkages, use of data in educational settings, personalization and recommendation services in astronomical data archives, intelligent retrieval of data, autonomous classification of objects, quantitative scoring of astronomical classifications of new objects, discovery of "interesting" objects and new classes of objects, information retrieval metrics on archive queries (precision and recall metrics), decision support for new observations and instrument-steering, query-by-example functionality in astronomical databases, and development of an astronomical genome. The latter will lead to the specification of key "genes" that define each class of astronomical object (e.g., principal components in multi-dimensional parameter spaces of multi-wavelength data). Isolating these "genes" will enable the construction of robust classifiers and parameter predictors (such as photometric redshifts) for massive petascale sky surveys, such as the LSST.

### The Astroinformatics Research Agenda – Requirements and Data Challenges

The new informatics approach to science covers many aspects of data-intensive research. We now summarize several instances of informatics research requirements, thereby specifying key features of an Astroinformatics research agenda and illustrating directions for Astroinformatics research and development. We strongly recommend that funding agencies adopt research plans in Astroinformatics, in the next decade, along the lines of these successful informatics programs.

An NIH grant solicitation in the field of health informatics identified a suite of **general themes in informatics research** (from the NIH website [35]):

- *Information and knowledge processing, natural language processing, information extraction, integration of data from heterogeneous sources, event detection, feature recognition.*
- *Tools for analyzing and/or storing very large datasets, data supporting ongoing experiments, and other data used in scientific research.*
- *Knowledge representation, including vocabularies, ontologies, simulations, and virtual reality.*
- *Linkage of experimental and model results to benefit research.*
- *Innovative uses of information technology in science applications, including decision support, error reduction, outcomes analysis, and information at the point of end-use.*
- *Efficient management and utilization of information and data, including knowledge acquisition and management, process modeling, data mining, acquisition and dissemination, novel visual presentations, and stewardship of large-scale data repositories and archives.*
- *Human-machine interaction, including interface design, use and understanding of science discipline-specific information, intelligent agents, information needs and uses.*
- *High-performance computing and communications relating to scientific applications, including efficient machine-machine interfaces, transmission and storage, real-time decision support.*
- *Innovative uses of information technology to enhance learning, retention and understanding of science discipline-specific information.*

The NASA Earth Science Technology Office has established goals for its Applied Information Systems Technology and REASoN programs [36] aimed at maximizing the science return from Earth Science programs. A subset of those goals includes: (a) provide data products, information systems and services capabilities, and/or advanced data systems technologies integrated into the project, to address strategic needs in science research, applications, and education; (b) increase the accessibility and utility of science data; (c) enable new observation measurements and information products; (d) increase science data value by responding to dynamic science using autonomous technologies; and (e) improve access, storage and delivery of large data volumes.

The report "Towards 2020 Science" identifies many research components of the informatics approach to science, in broad categories: (a) *intelligent interaction and information discovery* (managing the data explosion, adaptive organization and placement of data and computation, tools for data analysis, integrated symbolic computation, data mining and analysis, data structures for data visualization, empowering data-intensive scientists), and (b) *transforming scientific communication* (data display, dynamic delivery, links to deep data in scientific journals, digital discovery, computational thinking, and engaging the public in science) [4].

There are three other reports that contribute significant lists of requirements for informatics ("big data") research in astronomy. A recent 2008 DOE workshop on "Mathematics for Analysis of Petascale Data" produced a detailed report on research challenges and requirements to meet the needs of petascale data science [37], with inputs from astronomy [38]. Summarizing the major data-intensive research requirements of the LSST project, two additional (astronomy-specific) sources of Astroinformatics research requirements are the LSST Petascale Data R&D Challenges [39] and Data Management & Mining Challenges [40] documents, which evolved out of the original 1998 draft that was the first white paper on the topic of "big data" in astronomy.

**Existing Programs**

There are several existing and emerging programs in astronomical information sciences (or related informatics and data science disciplines) at various universities. The GMU graduate CSI (Computational Science and Informatics) program has existed since 1992 (covering many science disciplines, including astronomy). GMU's new undergraduate CDS (Computational and Data Sciences) degree program is now in its second year – half of the faculty is astronomers. Similar programs include: the Vanderbilt Initiative in Data-Intensive Astrophysics (VIDA); Caltech's Center for Advanced Computation and Research (CACR); POCA (Partnership in Observational and Computational Astronomy) at SCSU and Clemson; Purdue's Discovery Informatics program; the emerging joint programs between CS and astronomy departments at Notre Dame, and similarly at U. Michigan; the new e-Science Institute at U. Washington; Cornell's new DISCOVER data-driven science program; the "Tetherless World" Web Science program at RPI; and the strong astroinformatics program at Edinburgh University. Clearly there is an emerging academic focus on the astronomical information science disciplines.

Parallel to this, within the AGU professional society, a focus group on Earth and Space Science Informatics (ESSI) was formed in 2005. Since then, the numbers of ESSI presentations at their annual fall meetings have been substantial: 295 papers in Fall 2005, 345 (Fall 2006), 318 (Fall 2007), and 461 (Fall 2008). The interest within the Earth and Space Science communities on informatics-related scientific research problems is truly remarkable and robust. The formation of a similar focus group within the astronomical research community is timely, if not past due. See Thomas Loredo's Astronomical Information Sciences companion paper for more ESSI details.

**Recommendations**

This paper has outlined a research agenda for Astroinformatics and advocates for national support in the development of programs to advance the field as a viable formal sub-discipline of astronomy research and education, which builds on the VO experience, taking us from e-Resources through e-Science into the wealth of new knowledge to be discovered in massive data collections from sky surveys of the future. Our goal is to raise the productivity of science and to reap the maximum knowledge discovery potential from the vast astronomical data resources coming in the next decade. Astroinformatics borrows heavily from concepts in the fields of bioinformatics and geoinformatics, so that the gaps between these otherwise disconnected disciplines narrow, leading naturally to collaborations and cross-fertilization of ideas. As bioinformatics represents a modern paradigm for research in the biological sciences and is necessarily more data-oriented than computational biology, so also does astroinformatics represent a new mode of data-intensive scientific research in astronomy that is cognizant of and responsive to the astronomical data flood. The development of a shared common set of tools, methods, models, and approaches for data-intensive astronomy will enable collaborative research and data re-use (both in the research environment and in the classroom).

We specifically make the following recommendations to the State of the Profession Group:

1. To colleges and universities: support and sustain research programs, degree programs, and graduate training programs in Astroinformatics. Provide career development opportunities, support, and recognition for cross-disciplinary faculty.
2. To national labs and research centers: establish cyberinfrastructure (including databases and data-intensive computing platforms) that enable informatics research. Organize conferences and workshops that focus on the data science challenges within specific areas of astronomical research.
3. To funding agencies: Develop plans and programs that fund Astroinformatics research (see sample research program items from other agencies in preceding pages). Instruct peer reviewers in the unique evaluation metrics that apply to these new multi-disciplinary research proposals, which bridge both data sciences and astronomical sciences. Develop programs that fund the development of curricula and educational programs at the intersection of astronomical and data sciences. Support collaborative research with industry that utilizes emerging technologies for data-intensive science.
4. To professional societies: Promote conferences and IAU activities that address the development of the Astroinformatics sub-discipline. Initiate contacts with other disciplines' professional societies in order to explore common informatics research directions that can address problems in Astroinformatics. Establish an AAS Committee on Astroinformatics, which would provide a natural focal point for study and evaluation of the field, and would report to the AAS Council and Executive Committee regarding the status of the astronomical information sciences career path and the research productivity of its community of scientists.

## Co-Signers and Contributing Authors:

## Astronomers:

Kirk D. Borne, George Mason University
Alberto Accomazzi, Harvard-Smithsonian Center for Astrophysics
Joshua Bloom, University of California, Berkeley
Robert Brunner, University of Illinois at Urbana-Champaign
Douglas Burke, Harvard-Smithsonian Center for Astrophysics
Nathaniel Butler, University of California, Berkeley
David F. Chernoff, Cornell University
Brian Connolly, University of Pennsylvania
Andrew Connolly, University of Washington
Alanna Connors, Eureka Scientific
Curt Cutler, California Institute of Technology
Shantanu Desai, University of Illinois at Urbana-Champaign
George Djorgovski, California Institute of Technology
Eric Feigelson, Penn State University
L. Samuel Finn, Penn State University
Peter Freeman, Carnegie Mellon University
Matthew Graham, California Institute of Technology
Norman Gray, University of Leicester
Carlo Graziani, University of Chicago
Edward F. Guinan, Villanova University
Jon Hakkila, College of Charleston
Suzanne Jacoby, LSST Corporation
William Jefferys, University of Vermont and University of Texas at Austin
Vinay Kashyap, Harvard-Smithsonian Center for Astrophysics
Brandon Kelly, Harvard-Smithsonian Center for Astrophysics
Kevin Knuth, University at Albany
Donald Q. Lamb, University of Chicago
Hyunsook Lee, Harvard-Smithsonian Center for Astrophysics
Thomas Loredo, Cornell University
Ashish Mahabal, California Institute of Technology
Mario Mateo, University of Michigan Ann Arbor
Bruce McCollum, California Institute of Technology
August Muench, Harvard College Observatory
Misha (Meyer) Pesenson, California Institute of Technology
Vahe Petrosian, Standford University
Frank Primini, Harvard-Smithsonian Center for Astrophysics
Pavlos Protopapas, Harvard University
Andy Ptak, Johns Hopkins University
Jean Quashnock, Carthage College & University of Chicago
M. Jordan Raddick, Johns Hopkins University
Graca Rocha, Jet Propulsion Laboratory, California Institute of Technology

Nicholas Ross, Penn State University
Lee Rottler, IPAC/California Institute of Technology
Jeffrey Scargle, NASA Ames Research Center
Aneta Siemiginowska, Harvard-Smithsonian Center for Astrophysics
Inseok Song, University of Georgia
Alex Szalay, Johns Hopkins University
J. Anthony Tyson, University of California, Davis
Tom Vestrand, Los Alamos National Laboratory
John Wallin, George Mason University
Ben Wandelt, University of Illinois at Urbana-Champaign
Ira M. Wasserman, Cornell University
Michael Way, NASA Ames Research Center
Martin Weinberg, University of Massachusetts Amherst
Andreas Zezas, Harvard-Smithsonian Center for Astrophysics

## Information scientists:

Ethan Anderes, University of California, Davis
Jogesh Babu, Penn State University
Jacek Becla, SLAC National Accelerator Laboratory
James Berger, Duke University and Statistical and Applied Mathematical Sciences Inst.
Peter J. Bickel, University of California, Berkeley
Merlise Clyde, Duke University
Ian Davidson, University of California, Davis
David van Dyk, University of California, Irvine
Timothy Eastman, Wyle Information Systems
Bradley Efron, Stanford University
Chris Genovese, Carnegie Mellon University
Alexander Gray, Georgia Institute of Technology
Woncheol Jang, University of Georgia
Eric D. Kolaczyk, Boston University
Jeremy Kubica, Google
Ji Meng Loh, Columbia University
Xiao-Li Meng, Harvard University
Andrew Moore, Carnegie Mellon University
Robin Morris, Universities Space Research Association
Taeyoung Park, University of Pittsburgh
Rob Pike, Google
John Rice, University of California, Berkeley
Joseph Richards, Carnegie Mellon University
David Ruppert, Cornell University
Naoki Saito, University of California, Davis
Chad Schafer, Carnegie Mellon University
Philip B. Stark, University of California, Berkeley
Michael Stein, University of Chicago
Jiayang Sun, Case Western Reserve University
Daniel Wang, SLAC National Accelerator Laboratory

Xiao Wang, University of Maryland, Baltimore County
Larry Wasserman, Carnegie Mellon University
Edward J. Wegman, George Mason University
Rebecca Willett, Duke University
Robert Wolpert, Duke University
Michael Woodroofe, University of Michigan